\documentclass{llncs}
\usepackage{llncsdoc}
\usepackage{graphicx}
\usepackage{cite}
\usepackage{listings}
\usepackage{xcolor}
\usepackage{caption}
\usepackage{subcaption}
\usepackage{tikz}
\usepackage{amsfonts}
\usepackage{amssymb}
\usepackage{formal-grammar}
\usepackage{mathtools}
\usepackage{isabelle-listings}
\usepackage{graphicx}
\usepackage[dvips,bookmarksopen]{hyperref}
\usepackage{textcomp}
\usepackage{textgreek}
\usepackage{breakurl}
\usepackage{orcidlink}
\usepackage[bottom]{footmisc}

\usetikzlibrary{shapes,arrows}
\usetikzlibrary{arrows.meta, 
                bending,     
               }

\hypersetup{
	pdftitle={A Formal CHERI-C Semantics for Verification},    
	colorlinks=true,                          
	linkcolor=black,                          
	citecolor=blue,                           
	urlcolor=blue                             
}

\definecolor{codegray}{rgb}{0.5,0.5,0.5}

\lstdefinestyle{ccode}{
    keywordstyle=\bfseries,
    numberstyle=\tiny\color{codegray},
    basicstyle=\ttfamily\footnotesize,
    breaklines=true,
    captionpos=b,
    keepspaces=true,
    numbers=left,
    numbersep=7pt,
    showspaces=false,
    showstringspaces=false,
    showtabs=false,
    tabsize=2
}

\lstdefinestyle{isacode}{
    aboveskip=10pt,
    belowskip=10pt,
    language={Isabelle},
    breaklines=true,
    breakatwhitespace=true,
    mathescape=true,
}

\newcommand{\Coloneqq}{\ \triangleq\ }

\renewcommand{\gralt}[0]{\ |\ }
\newcommand{\xMapsto}[2][]{\ext@arrow 0599{\Mapstofill@}{#1}{#2}}
\makeatletter\@twosidefalse\@mparswitchfalse\makeatother
\begin{document}
\title{A Formal CHERI-C Semantics for Verification}
%
\author{Seung Hoon Park~\orcidlink{0000-0001-7165-6857} \and Rekha Pai~\orcidlink{0000-0002-5964-8819} \and Tom Melham~\orcidlink{0000-0002-2462-2782}}
\institute{Department of Computer Science, University of Oxford, Oxford, UK
\email{\{seunghoon.park,rekha.pai,tom.melham\}@cs.ox.ac.uk}}

\maketitle \thispagestyle{empty}

\begin{abstract}
CHERI-C extends the C programming language by adding \textit{hardware
capabilities}, ensuring a certain degree of memory safety while remaining efficient.
Capabilities can also be employed for higher-level security measures, such as
software compartmentalization, that have to be used correctly to achieve the
desired security guarantees. As the extension changes the semantics of C,
new theories and tooling are required to reason about CHERI-C code
and verify correctness.
In this work, we present a formal memory model that provides a memory
semantics for CHERI-C programs. We present a generalised theory with rich
properties suitable for verification and potentially other types of
analyses. Our theory is backed by an Isabelle/HOL formalisation that also
generates an OCaml executable instance of the memory model. The verified and
extracted code is then used to instantiate the parametric
\textit{Gillian}
program analysis framework, with which we can perform concrete execution of
CHERI-C programs. The tool can run a CHERI-C test suite, demonstrating the
correctness of our tool, and catch a good class of safety violations that the
CHERI hardware might miss.
\end{abstract}

\section{Introduction}
Despite having been developed more than 40 years ago, C remains
a widely used programming language owing to its efficiency, portability, and suitability for low-level
systems code. 
The language's lack of inherent memory safety, however, has
been the source of many serious issues\cite{chisnall_2015}. While there
have been significant efforts aimed at vulnerability mitigation, memory safety
issues
remain widespread, with a recent study stating that 70\% of security
vulnerabilities are caused by memory safety issues\cite{miller_2019}.

The Capability Hardware Enhanced RISC Instructions (CHERI) project offers
an alternative model that provides better memory safety~\cite{cheri2014}. Its main
features include a new machine representation of C pointers called
\textit{capabilities} and extensions to existing Instruction Set Architectures
(ISA) that enable the secure manipulation of capabilities.
Capabilities are in essence memory addresses bound to
additional  safety-related metadata, such as access permissions and bounds on
the memory locations that can be accessed. As the hardware performs
the safety checks on capabilities, legacy
C programs compiled and run on CHERI architecture, i.e. CHERI-C code,
acquire hardware-ensured spatial memory safety, while retaining
efficiency. Porting code from one language to another generally requires
significant efforts. But porting C codes to CHERI-C requires little, if any, changes
to the original code to ensure the code runs on CHERI hardware~\cite{watson_laurie_richardson_2021, richardson_2022}.

In 2019, the UK announced its \textit{Digital Security
by Design} programme with \pounds 190 million of funding  distributed over more than 26 research projects and 5 industrial demonstrators
~\cite{DigitalSecurityDesign} to `radically update the foundation of
our insecure digital computing infrastructure, by demonstrating that mainstream
processor technology \dots\ can be updated to include new security technologies
based on the CHERI Architecture'~\cite{DepartmentComputerscienceb}.  A
cornerstone of the programme is Morello~\cite{DepartmentComputerScience}, a
CHERI-enabled prototype developed by Arm.

\begin{figure}[t]
    \centering
    \begin{subfigure}[b]{0.45\textwidth}
        \centering
        \includegraphics[scale=0.65]{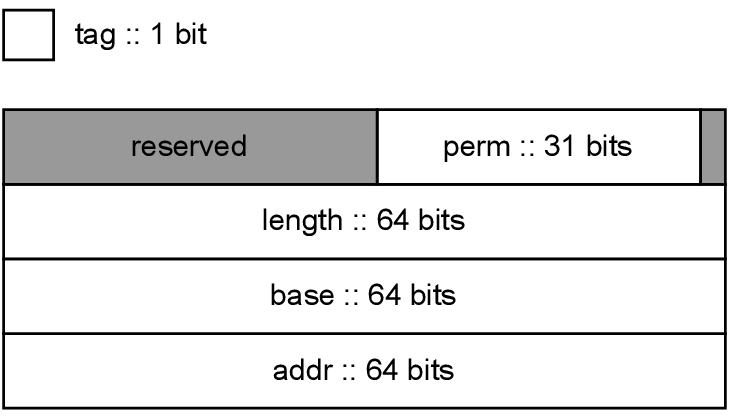}
        \caption{CHERI-256 Capability Layout}
        \label{fig:cap256}
    \end{subfigure}
    \hfill
    \begin{subfigure}[b]{0.45\textwidth}
        \centering
        \includegraphics[scale=0.65]{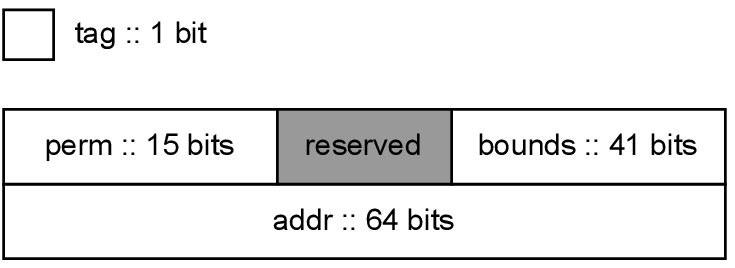}
        \caption{CHERI-128 Capability Layout}
        \label{fig:cap128}
    \end{subfigure}
    \caption{Simplified CHERI Capability Layouts}
    \label{fig:cap}
\end{figure}

Over the several years that lead to the realisation of
Morello, there were several design revisions made to the
hardware; examples are depicted in Fig.~\ref{fig:cap}.
The refined designs used methods for compression of bounds that
reduced cache footprints and improved overall performance while minimising
incompatibility.
Morello uses a very similar design to the compressed scheme for capabilities
depicted in Fig.~\ref{fig:cap128},
with the overall bit-representation of the layout differing slightly. Future
capability designs may possibly incorporate a different bit-representation design,
provided there are improvements in performance or compatibility. Due to the
ever-changing design of capability bit-representations, it seems best to have an
\textit{abstract} representation of capabilities, so that CHERI-based verification
tools can remain modular.

Checking for memory safety issues of legacy C code can, of course, be achieved using existing
analysis tools for C, but there are new problems that arise when such code is
run on CHERI hardware. Because the pointer and memory representations
are fundamentally different in a CHERI architecture, there are non-trivial
differences in the semantics between C and CHERI-C.

To illustrate this point,
consider the C code in Listing \ref{lst:c_example}.
This code segment performs \verb|memcpy| twice: once from \verb|a| to \verb|b|,
where pointers/capabilities are stored misaligned in \verb|b|, then from
\verb|b|
to \verb|c|, where pointers/capabilities are stored correctly again in \verb|c|.
In standard C, there are no problems accessing the pointer stored in \verb|c|.
But in CHERI-C, misaligned capabilities in memory are invalidated. That means
the address and meta-data of the misaligned capabilities are accessible, but
such
capabilities can no longer be dereferenced \cite{watson_cc_2019}. While \verb|c| will
contain the same capability value as that of \verb|a|, the capability stored in
\verb|c| is invalidated. Thus, the last line will trigger an `invalid tag'
exception when the code is executed on ARM Morello and other CHERI-based
machines.
\begin{lstlisting}[language=C, caption=C code example, label={lst:c_example}]
#include <stdlib.h>
#include <string.h>
void main(void) {
    int *n = calloc(sizeof(int), 1);
    int **a = malloc(sizeof(int *));
    *a = n;
    int **b = malloc(sizeof(int *) * 2);
    int **c = malloc(sizeof(int *));
    memcpy((char *) b + 1, a, sizeof(int *));
    memcpy(c, (char *) b + 1, sizeof(int *));
    int x = **c;
}
\end{lstlisting}

Of course, existing C analysis tools cannot catch these cases, as such
tools are not only unaware of the changes in the semantics that
capabilities bring, but also the code is not problematic in conventional C.
Moreover, while CHERI ensures spatial safety by the hardware, CHERI is
still incapable of catching temporal safety violations, such as Use After
Free (UAF) violations.  There exists work that attempt to address temporal
safety~\cite{filardo_2020, chisnall_2022, mte_2021}, but they are either a
software-implemented solution \cite{filardo_2020}, where overall
performance is inevitably affected, or ongoing work \cite{mte_2021}. There
is, therefore, a need for program analysis tools  that correctly
integrate the semantics of CHERI-C.

To the best of our knowledge, there is no prior work on formalising a
CHERI-C memory model. The Cerberus C work \cite{memarian_2019} is primarily
designed to capture pointer provenance of C programs and uses CHERI-C as a
reference for pointer provenance, but the tool lacks a formal CHERI-C
memory model. ESBMC is a verification tool that supports CHERI-C code
\cite{brausse_2022}. But support for tagged memory does not yet exist;
ESBMC would not be able to catch the `invalid tag' exception in the code
in Listing \ref{lst:c_example}. Furthermore, ESBMC's memory model is not
formally verified. Users of ESBMC must trust that the implementation of
the memory model and its underlying theory are correct. SAIL formalisations
for each CHERI architectures exist\cite{cheri_mips_github, cheri_riscv_github,
cheri_morello_github}, but they only capture the low-level semantics of the
architecture and not high-level C constructs such as
allocation.

In this paper, we introduce a formal CHERI-C memory model that captures the
memory semantics of the CHERI-C language. In Sect.~\ref{sec:ccmm}, We
formalise the memory and its
operations and prove essential properties that provide correctness
guarantees. We provide a rigorous logical formalisation
of the CHERI-C
memory model in Isabelle/HOL\cite{nipkow_2002} (in Sect.~\ref{sub:isabelle})
and use the code generation
feature to generate a verified OCaml instance of the memory model
\cite{haftmann_2021}. We then show, in Sect.~\ref{sub:Gillian}, the practical
aspects of this work by providing the memory model to, and thereby
instantiating, Gillian \cite{fragoso_2020}, a general, parametric
verification framework that supports concrete and symbolic execution and
verification based on separation logic, backed by rich correctness
properties. In Sect.~\ref{sec:expt}, we demonstrate that the tool can capture
the semantics of CHERI-C programs correctly. A discussion on the
existing works can be found in Sect.~\ref{sec:related} while
Sect.~\ref{sec:concl} concludes this paper mentioning possible future
directions. We first start with an introduction to the CHERI
architecture.
%
\section{CHERI}\label{sec:cheri}
CHERI extends a conventional ISA by introducing \emph{capabilities} which are
essentially pointers that come along with metadata to restrict
memory access. The ISA now has
additional hardware instructions and exceptions that operate over capabilities.
Register sets are extended to include capability registers,
instructions are added that reference the capability registers,
and custom hardware exceptions are added to
block operations that would violate memory safety.
Designs of CHERI capabilities have refined over the past several years
and have been incorporated in several existing architectures, such
as MIPS and RISC-V~\cite{watson_et_al_2020}. All CHERI-extended ISAs have been
formally defined using the SAIL specification language, in which the logic of
machine instructions and memory layout have been defined formally in a
first-order language~\cite{armstrong_2019}.

Regardless of the layout, CHERI
capabilities include three important types of high-level information, in
addition to a 64-bit address:
\begin{itemize}
    \item \textbf{Permissions.} Permissions state what kind of operations a
          capability can perform. Loading from memory and storing to memory are
          examples of permissions a capability may possess.
    \item \textbf{Bounds.} Bounds stipulate the memory region that the address
part of a capability can
          reference. The lower bound stipulates the lowest address that
          a capability may access, and the upper bound stipulates the highest
          address.
    \item \textbf{Tag.} Stored separately from the other components of a
          capability, the tag states the validity of the capability it is attached to. Capabilities
          with invalid tags can hold data but cannot be dereferenced. Attempts
          to forge capabilities out of thin air result in a tag-invalidated
          capability.
\end{itemize}

Fig.~\ref{fig:cap256} show a 256-bit representation of a capability, which
was one of the earlier designs. The lower and upper bounds are represented using
the base and length fields. Here, the lower bound is the address stated by the
base field, and the upper bound is the address in the base field plus the length
field. Permissions and other metadata are stored in the remaining fields as a
bit vector.
The capability's tag bit exists separately from the capability. Tag bits are, in
practice, stored separately from the main memory where capabilities reside, so
users cannot manipulate the tag bits of capabilities stored in memory.
Furthermore, overwriting capabilities stored in memory with non-capability
values invalidates their tag bits, which ensures
capabilities cannot be forged out of thin air.

This representation, in theory,
exercises a high level of compatibility with existing C code. But performance,
particularly with regards to caching, is reduced due to the size of the
capability representation~\cite{woodruff_2019}. Refined designs ultimately resulted in a
capability that utilises a floating-point-based lossy compression technique
on the bounds \cite{woodruff_2019}, such as the one depicted in Fig.~\ref{fig:cap128}. In many cases, the upper bits of the
address fields are most likely to overlap with those of the lower and upper
bounds. Knowing this, bounds can be compressed by having the upper bits of
their fields depend on that of the address, which means only the lower bits
need to be stored.

The lossy compression of bounds may result in some incompatibility. Bounds
may no longer be represented exactly, and changes in the address field may
result in an unintentional change in the bounds.  Nonetheless, such
representations give an acceptable level of compatibility, provided aggressive
pointer arithmetic optimisations are avoided. The Morello processor incorporates a similar
compression-based design in its architecture, though sizes of each field
differ~\cite{morello_2022}.

The added capability-aware instructions operate over capabilities. Conventional
load and store operations are extended to first check that the tag,
permissions, and bounds of the capability are all valid. Violations result in triggering a
capability-related hardware exception. There are additional operations to
access or change the tag, permissions, and bounds. To ensure spatial memory
safety, these operations can, at most, make the conditions for execution more
restrictive; they cannot grant that which was not previously available. For instance, one cannot
lower the lower bound of a capability to access a region that was inaccessible
before, or grant a store permission that was unset beforehand. Because of how
tags work for capabilities stored in memory, one cannot grant capabilities
larger bounds or more permissions by manipulating the memory---attempting this results
in tag invalidation.

Library support for CHERI has grown over the past few years. In particular,
a software stack for CHERI-C that utilises a custom Clang compiler now exists
\cite{watson_cc_2019}. Users can compile their program either in `purecap' mode,
where all pointers in programs are replaced with capabilities, or in `hybrid'
mode, where both pointers and capabilities co-exist within the program.
Because operations that change the fields of a capability does not generally
exist in standard C, Clang incorporates additional CHERI libraries of
operations that users may use to access or mutate capabilities.

\section{CHERI-C Memory Model} \label{sec:ccmm}
Incorporating hardware-enabled spatial safety requires  significant changes
to the C memory model. Pointer designs must be extended to
incorporate bounds, metadata, and the out-of-band tag bit. The memory, i.e.
heap, must also be able to distinguish the main memory and the tagged memory.
Operations with respect to the heap must also be defined such that tag
preservation and invalidation are incorporated appropriately.

In this section, we provide a generalised theory for the CHERI-C memory model.
We identify the type and value system used by the memory model. We then
define the heap and the core memory operations. Finally, we state some essential
properties of the heap and the operations that (1) characterises the semantics
and (2) states what types of verification or analyses could be supported. We
make the assumption that we work on a `purecap' environment, where all
pointers have been replaced with capabilities.
\subsection{Design}
The CHERI-C memory model is inspired by that of
CompCert\cite{leroy_2012}. The beauty of CompCert is that it is a verified
C compiler. The internal components, which include the block-offset based
memory model, are formalised in a theorem prover, with many of its essential
properties verified. Using CompCert's memory model as a basis, we design the
CHERI-C memory model by providing extensions to ensure the modelling
of correct semantics and the capture of safety violations:
\begin{itemize}
    \item \textbf{Capability Values.} In addition to the standard primitive
          types, we incorporate abstract capabilities as values.
          We also incorporate capability fragments to provide semantics to
          higher-level memory actions like \verb|memcpy|, which should preserve
          tags if copied correctly and invalidate otherwise \cite{watson_cc_2019}.
    \item \textbf{Extended Operations.} Basic memory actions such as \verb|load|
          and \verb|store| now work on capabilities and will trigger the
          correct capability-related exception when required.
    \item \textbf{Tagged Memory.} Tags in memory are stored separately from the
          main heap, as could be seen by the formal CHERI-MIPS SAIL model
          \cite{cheri_mips_github}. So we provide a separate mapping for
          tagged memory for storing capability tags.
    \item \textbf{Freed Regions.} The standard CompCert memory model can mark
          which memory regions are valid but lacks the ability to distinguish
          which regions are marked as `Freed'. We incorporate freed regions 
          as a means to catch temporal safety violations.
\end{itemize}

\subsection{Type and Value System}

\begin{grammar}[CHERI-C Types and Values][b][fig:cc_types_values]
    \firstcasesubtil{$\tau$}{U8_\tau \gralt S8_\tau \gralt ... \gralt U64_\tau
    \gralt S64_\tau \gralt Cap_\tau}{}
    \firstcasesubtil{\textit{MCap}}{\mathcal{B} \times \mathbb{Z} \times md}
    {}
    \firstcasesubtil{\textit{Cap}}{MCap \times \mathbb{B}}{}
    \firstcasesubtil{$\mathcal{V}_{\mathcal{C}}$}{U8_\mathcal{V}\ ::\ 8\ bits \gralt ...}{}
    \otherform{S64_\mathcal{V}\ ::\ 64\ sbits}{}
    \otherform{Cap_\mathcal{V}\ ::\ Cap}{}
    \otherform{CapF_\mathcal{V}\ ::\ Cap \times \mathbb{N}}{}
    \otherform{Undef}{}
    \firstcasesubtil{$\mathcal{V}_\mathcal{M}$}{Byte\ ::\ 8\ bits}{}
    \otherform{MCapF\ ::\ MCap \times \mathbb{N}}{}
\end{grammar}

Figure \ref{fig:cc_types_values} shows the formalisation of CHERI-C types and values.
Types $\tau$ are analogous to chunks in CompCert terms. Types
comprise primitive types (e.g.~U8$_\tau$, S64$_\tau$, etc.)
and a capability type $Cap_\tau$. We define a function $|\cdot|\ :\ \tau
\rightarrow \mathbb{N}$ that
returns, in terms of bytes, the size of the type. For $Cap_\tau$, the value is not
fixed but requires that it must be divisible by 16. This requirement
allows capabilities with 128- and 256-bit representations to have
a valid size.

$MCap$ represents a \textit{memory capability} value and is
represented
as a tuple $(b, i, m)$, which comprises the block identifier $b \in \mathcal{B}$,
offset $i \in \mathbb{Z}$, and metadata $m \in md$, where $md$
represents the bounds and permissions.
Here, $\mathcal{B}$ must be a countable set.
 Offsets are represented as integers, as
CHERI allows out-of-bounds addresses, where the address may be lower than the lower
bound. Because capabilities stored in memory have
their tag bit stored elsewhere, we make the distinction between memory
capabilities
and \textit{tagged capabilities}, $\mathit{Cap}$,  which is a capability $((b, i, m),
t)$
that contains the tag bit $t \in \mathbb{B}$.

Unlike those of CompCert, CHERI-C values $\mathcal{V}_{\mathcal{C}}$
are given type distinctions to ensure: (1) types can be inferred directly, and
(2) they contain the correct values at all times. From a practical standpoint,
this ensures that the proof of correctness of memory
operations can be simplified, and bounded arithmetic operations can be implemented correctly.
Capability values $Cap_\mathcal{V}$ and capability
fragment values $CapF_\mathcal{V}$ also exist as values. Provided some capability 
value $C \in Cap_{\mathcal{V}}$, capability fragment
values $C_n \in CapF_\mathcal{V}$ correspond to the $n$-th byte of the capability
$C$. For both cases,
instead of fixing their representation concretely, we represent
them abstractly using a tuple. This representation ensures that conversion to a
compressed representation could be achieved when needed while avoiding the need
to fix to one particular bit representation. Furthermore,
this approach provides a reasonable way to correctly define \texttt{memcpy},
where capability tags must be preserved if possible.
While capability fragments are extended structures of capabilities, operations
that can be performed on capability fragments are limited. Finally, we
have $\mathit{Undef}$, which represents invalid values. These values may appear when, for
example, the user calls \texttt{malloc} and immediately tries to load the
undefined contents. The idea behind incorporating capability fragments values is
heavily inspired by the work from \cite{krebbers_2014}.

Because values are given a type distinction, identifying the types of values is
straightforward. For capability fragments, we have two choices: they may either be
a $U8_\tau$ or $S8_\tau$ type. Capability fragments are essentially bytes, so
operations over capability fragments can be treated as if they were
a $U8_\tau$ or $S8_\tau$ type. Since $\mathit{Undef}$ does not correspond to
a valid value, it is not assigned a type.

\begin{grammar}[CHERI-C Errors][thp][fig:cc_errors]
    \firstcasesubtil{\text{CapErr}}{\text{TagViolation} \gralt \text{PermitLoadViolation} \gralt \ldots}{}
    \firstcasesubtil{\text{LogicErr}}{\text{UseAfterFree} \gralt \text{MissingResource} \gralt \dots}{}
    \firstcasesubtil{\text{Err}}{\text{CapErr} \gralt \text{LogicErr}}{}
    \firstcasesubtil{$\mathcal{R}\ \rho$}{\mathcal{S}ucc\ \rho}{}
    \otherform{\mathcal{F}ail\ \text{Err}}{}
\end{grammar}

Memory operations, such as \texttt{load} and \texttt{store}, are defined so
that, upon failure, the operation returns the type of error that lead to the
failure. In general, partial functions, or function using the option type,
can model function failure but cannot state what caused the failure. As such,
the operations use the return type $\mathcal{R}\ \rho$, where $\rho$ is
a generic return type. For CHERI-C, we make the distinction between errors
caused by capabilities, denoted by CapErr, and errors caused by the language,
denoted by LogicErr. Figure \ref{fig:cc_errors} depicts the formalised Errors
system used by the memory model.

\subsection{Memory}

We now formalise the memory. We use CompCert's approach of using a union type
$\mathcal{V}_\mathcal{M}$ that can represent either a byte or a byte fragment of
a memory capability. Then it is possible to create a memory mapping $\mathbb{N}
\rightharpoonup \mathcal{V}_\mathcal{M}$.\footnote{The notation
$\rightharpoonup$ denotes a partial map. Offsets in heaps are $\mathbb{N}$,
whereas offsets stored in capabilities are $\mathbb{Z}$. Operations check
whether the offsets are in bounds, which requires offsets to be non-negative.
This means valid offset values can be converted from $\mathbb{Z}$ to $\mathbb{N}$
without issues.}
We also create a separate mapping of type $\mathbb{N} \rightharpoonup \mathbb{B}$ for
tagged memory. When the user attempts to store a capability, it
will be converted into a memory capability and then stored in the memory mapping.
Separately, the tag bit will be stored in the tagged memory. When
the tag bit is stored, adjustments are made to ensure tags are only stored in
capability-size-aligned offsets.

To ensure we can catch temporal safety violations, we need to be able to
make distinctions between blocks that are freed and blocks that are valid. One way to
encode this is as follows: a block $b$ may point to either a freed location
(i.e. $b \mapsto \varnothing$), or point to the pair of maps we defined earlier. The idea
is that if a block identifier points to a freed block, attempts to load such a block will
 trigger a `Use After Free' violation and would otherwise point to a valid
mapping
pair. Ultimately, the heap has the following form:
$$\mathcal{H} : \mathcal{B} \rightharpoonup ((\mathbb{N} \rightharpoonup
\mathcal{V}_\mathcal{M}) \times (\mathbb{N} \rightharpoonup \mathbb{B}))_{\varnothing}$$

\subsection{Operations}

We define the core memory operations, or \emph{actions}, of the memory model.
We use
the same result type $\mathcal{R}$ given in Fig.~\ref{fig:cc_errors} instead
of using a partial function to give the type of error, should the operation fail.

The memory actions $A_{\mathcal{C}} = \{\texttt{alloc}, \texttt{free},
\texttt{load}, \texttt{store}\}$ are given below with their
respective signatures:
\begin{itemize}
    \item $\texttt{alloc} :\ \mathcal{H} \rightarrow \mathbb{N} \rightarrow
          \mathcal{R}\ (\mathcal{H} \times Cap)$
    \item $\texttt{free } :\ \mathcal{H} \rightarrow Cap \rightarrow
          \mathcal{R}\ (\mathcal{H} \times Cap)$
    \item $\texttt{load } :\ \mathcal{H} \rightarrow Cap \rightarrow
          \tau \rightarrow \mathcal{R}\ (\mathcal{V}_{\mathcal{C}})$
    \item $\texttt{store} :\ \mathcal{H} \rightarrow Cap \rightarrow
          \mathcal{V}_{\mathcal{C}} \rightarrow \mathcal{R}\ (\mathcal{H})$
\end{itemize}

The function \texttt{alloc} $\mu\ n = \mathcal{S}ucc\ (\mu', c)$ takes a heap $\mu$ and size
$n$ input and produces a fresh capability $c$ and the updated heap $\mu'$ as output.
The bounds of $c$ are determined by $n$. In the case of compressed capabilities,
a sufficiently large $n$ \textit{may} result in the upper bound being larger than
what was requested. The capability $c$ is also given the appropriate permissions and a valid
tag bit. Like that of CompCert, \texttt{alloc} is designed to never fail,
provided that the countable set $\mathcal{B}$ has infinite elements.

The function \texttt{free} $\mu\ c = \mathcal{S}ucc\ (\mu', c')$ takes a heap $\mu$ and
capability $c = ((b, i, m), t)$ as input. Upon success, the operation will
return
the updated heap, where we now have $b \mapsto \varnothing$. The capability $c'$
is also updated such that the tag bit of $c$ is invalidated. This conforms to
the CHERI-C design stated in \cite{watson_cc_2019}. We note that $c$ should
also be a valid capability, that is---at the very least---the tag bit should be
set, and the offset should be within the capability bounds. The function \texttt{free}
may fail if the block is invalid or already freed, even if the capability
itself was valid. In such case, \texttt{free} returns a logical error.

The function \texttt{load} $\mu\ c\ t\ = \mathcal{S}ucc\ v$ takes a  heap $\mu$, capability
$c$ and type $t$ as input, where $t$ is the type the user wants to load. Upon
success, the operation will return the value $v$ from the memory, where $v$
has the corresponding type $t$.\footnote{For capability fragments, the corresponding
type may be either $U8_\tau$ or $S8_\tau$.} Before
\texttt{load} attempts to access the block provided by $c$, it first
checks that $c$ has sufficient permissions to load. We use the CHERI-MIPS
SAIL implementation of the CL[C] instruction \cite{watson_et_al_2020} for
the capability checks, implementing the extra checks provided that $t = Cap_\tau$.
Once the capability checks are done, the operation attempts to access
the blocks and the mappings, failing and returning the appropriate logical
error if they do not exist.

When accessing both the main memory and tagged
memory, there are a number of cases to consider. When loading primitive values,
it is important that the region about to be loaded is all of $\mathit{Byte}$ and not of
$\mathit{MCapF}$ type. Thus, before loading the values, we check whether the contiguous
region in memory are all of $\mathit{Byte}$ type. If this is not the case, \texttt{load}
will return $\mathit{Undef}$. For capability fragments, the cell in memory has to be
an $\mathit{MCapF}$. Finally for capabilities, not only do the contiguous cells have to
be of $\mathit{MCapF}$ type, but (1) they must  have the same memory capability value,
and (2) the fragment values must all be a sequence forming $\{0, 1, ..., |Cap_\tau| - 1\}$.
The idea is that even if the contiguous cells have the same memory capability
values, they do not form a valid capability if the fragments are not stored in order.
After all the checks, the tagged memory will be accessed, where the tag value
is retrieved.\footnote{The tagged memory does not need to be accessed if $c$
does not have a capability load permission. In such case, the loaded capability
will have an
invalidated tag.} The loaded memory capability and tag bit are then combined to form
a tagged capability, which \texttt{load} returns.

The function \texttt{store} $\mu\ c\ v\ = \mathcal{S}ucc\ \mu'$ takes a heap $\mu$, capability
$c$, and value $v$. Upon success, the operations will return the updated heap
$\mu'$. Like \texttt{load}, \texttt{store} performs the necessary capability
checks based on CHERI-MIPS' CS[C] instruction and attempts to access the blocks
and mappings afterwards, returning the appropriate exception upon failure.
For storing primitive values and capability fragment values, the main memory
mapping will simply be updated to contain the values, and the associated tagged
memories will be invalidated. For primitive values that
are not bytes, the values will be converted into a sequence of bytes, where
each byte in the list will be stored contiguously in memory. For a capability
fragment value, it will be stored in the cell as an $\mathit{MCapF}$ type, where the
tag value of the fragment will be stripped when storing in memory. Finally,
for capability values, the value will be split into a list comprising
$|Cap_\tau| - 1$ memory capability fragments, with the fragment value forming a
sequence $\{0, 1,.. |Cap_\tau| - 1\}$, and a tag bit. The main memory will store the
list of memory fragments contiguously, and the tagged memory will store the tag
value in the corresponding capability-aligned tagged memory.



\subsection{Properties}

In the previous section, we have articulated a formal CHERI-C memory model,
explaining how the heap is structured and how the operations are defined.
It is essential that the formalisation we provided is correct and is also
suitable for verification or other types of analyses. In this section, we  first discuss
the properties of the memory. We then discuss  the properties of the
operations themselves, primarily concerned with correctness.

When we observe the memory, it is important that we always work with a valid
one, i.e. the memory is \textit{well-formed}. In our formalisation, we require
that all tags in the tagged memory are stored in a capability-aligned location.
The well-formedness relation $\mathcal{W}^\mathcal{C}_{f}$ is defined as follows:
$$\mathcal{W}^\mathcal{C}_{f}(\mu) \equiv \forall b \in dom(\mu).\ b \mapsto (c, t) \longrightarrow
\forall x \in dom(t).\ x\ \text{mod}\ |\textit{Cap}_\tau| = 0$$

The well-formedness property must hold when the heap is initialised and when
memory operations mutate the heap. That is, provided $\mu_0$ is the initialised
heap where all mappings are empty, $\alpha \in A_{\mathcal{C}}$ is a memory action,
$v$ are the arguments of the memory operation $\alpha$ and $\mu'$ is one of the
return values denoting the updated heap, we have the following properties:
$$\mathcal{W}^\mathcal{C}_{f}(\mu_0)$$
$$\mathcal{W}^\mathcal{C}_{f}(\mu) \Longrightarrow
  \alpha\ \mu\ v = \mathcal{S}ucc\ \mu' \Longrightarrow
  \mathcal{W}^\mathcal{C}_{f}(\mu')$$

\noindent The two properties above ensure that the heap is well-formed throughout the
execution of the CHERI-C program.


For the correctness of the operations, we primarily consider soundness and
completeness:
\begin{itemize}
    \item If the inputs are valid for operation $\alpha \in A_\mathcal{C}$
          then the action should succeed.
    \item If the action $\alpha$ succeeds, the inputs provided to the operations
          are valid.
    \item If the inputs are invalid for the operation $\alpha$, then the
          action should fail and return the correct error.
\end{itemize}
The first and second points are simple soundness and completeness properties.
The third point is important in that the input may be problematic in many
ways. For example, the NULL capability has an invalid tag bit, invalid bounds,
and no permissions. The function \texttt{load} will fail if provided with the
NULL capability,
as it violates many of the checks. Because the SAIL specification states that
tags are always checked first, the error must be a TagViolation type.

Next, we need to ensure successive operations yield
the desired result. The primary properties to consider are the
\textit{good variable} laws \cite{leroy_2012}; examples of properties encoding
this law include \textit{load after allocation}, \textit{load after free}, and
\textit{load after store}.
It is worth mentioning there are some caveats. For example, the
\textit{load after store} case no longer guarantees that you will retrieve the
same value you stored, unlike CompCert's load after store property in
\cite{leroy_2012}, since the value that was stored and to be loaded again could
have been either a capability or capability fragment. In such cases, the tag
bit may become invalidated due to insufficient permissions on the capability, or
because storing capability fragments resulted in the tagged memory being cleared.
The solution is to divide the general property into a primitive value case and a
capability-related value case. Ultimately, the idea is to prove that the loaded
value is \textit{correct} rather than exact, i.e. capability-related values
when loaded with have the correct tag value.

Finally, we have properties suitable for verification. We note that the memory
$\mathcal{H}$ can be instantiated as a separation algebra by providing
the partial commutative monoid (PCM) $(\mathcal{H}, \uplus, \mu_0)$, where
$\uplus$ is the disjoint union of two heaps and $\mu_0$ is the empty
initialised heap. For tools that rely on using partial memories, it is also
imperative to show that the well-formedness property is compatible with memory
composition:
$$\mathcal{W}^\mathcal{C}_f(\mu_1 \uplus \mu_2) \Longrightarrow
  \mathcal{W}^\mathcal{C}_f(\mu_1) \land \mathcal{W}^\mathcal{C}_f(\mu_2)$$

\noindent We also note that the current heap design keeps track of \textit{negative}
resources \cite{maksimovic_2021}, which may potentially be useful for
incorrectness logic based verification\cite{ohearn_2019}.

\section{Application} \label{sec:app}

The overall memory model provided in Sect.~\ref{sec:ccmm} has been designed
to be applicable for verification tools. In this section, we
explain how we use the theory provided above to create a verified,
executable instance of the memory model. We then explain how this
executable model can be used to instantiate a tool called Gillian~\cite{fragoso_2020}. Using
the instantiated tool, we demonstrate the concrete
execution of CHERI-C programs with the desired behaviour.
\subsection{Isabelle/HOL}\label{sub:isabelle}
Isabelle/HOL is an interactive theorem prover based on classical Higher Order
Logic (HOL) \cite{nipkow_2002}. We use Isabelle/HOL to formalise the entirety of
the CHERI-C memory model discussed in Sect.~\ref{sec:ccmm}.
Types,
values, heap structure, etc. were implemented, memory operations were
defined, and properties relating to the heap and the operations were proven.
Memory capabilities, tagged capabilities, and capability fragments were
represented using records, a form of tuple with named fields. For code generation, we
instantiated the block type $\mathcal{B}$ to be $\mathbb{Z}$.
For showing that $\mathcal{H}$ is an instance of a separation
algebra, we use the \texttt{cancellative\_sep\_algebra} class \cite{klein_2012}
and prove that the heap model is an instance. This proof ultimately shows that
$\mathcal{H}$ forms a PCM. Proving that well-formedness is compatible with memory
composition is stated slightly differently. The
\texttt{cancellative\_sep\_algebra} class takes in a total operator $\cdot_t$
instead of a partial one and requires a `separation disjunction' binary
operator $\texttt{\#}$, which states disjointedness. Ultimately, the
compatibility
property can be given as: $$\mu_1\ \texttt{\#}\ \mu_2 \Longrightarrow
\mathcal{W}^\mathcal{C}_f (\mu_1 \cdot_t \mu_2) \Longrightarrow
\mathcal{W}^\mathcal{C}_f(\mu_1) \land \mathcal{W}^{\mathcal{C}}_f(\mu_2)$$
For partial mappings of the form $A \rightharpoonup B$, we use Isabelle/HOL's
finite mapping type \texttt{('a,'b)mapping} \cite{haftmann_2013}. To ensure we
obtain an OCaml executable instance of the memory model, we use the Containers
framework\cite{lochbihler_2013}, which generates a Red-Black Tree mapping
provided the abstract mapping in Isabelle/HOL. All definitions
in Isabelle were either defined to be code-generatable to begin with (i.e.
definitions should not comprise quantifiers or non-constructive constants
like the Hilbert choice operation $SOME$), or
code equations were provided and proven to ensure a sound code generation
\cite{haftmann_2021}. For bounded machine words, which is required for
formalising the primitive values, we use Isabelle/HOL's word type
\texttt{'a word}, where \texttt{'a} states the length of the word
\cite{beeren_2016}. Types like $\texttt{'a word}$, $\texttt{nat}$, $\texttt{int}$ and
$\texttt{string}$ were also transformed to use OCaml's Zarith and native string
library for efficiency \cite{haftmann_2021}.
\subsection{Gillian}\label{sub:Gillian}
Gillian is a high-level analysis framework, theoretically capable of
analysing a wide range of languages. The framework allows concrete and symbolic execution,
verification based on Separation Logic, and bi-abduction\cite{maksimovic_2021}. The crux of
the framework lies in its parametricity, where the tool can be instantiated by simply
providing a compiler front end and OCaml-based memory models of the language. So far, CompCert C and
JavaScript have both been instantiated for Gillian, giving birth to Gillian-C
and Gillian-JS.

The underlying theoretical foundation of Gillian has its
essential correctness properties like soundness and completeness already proven
\cite{fragoso_2020, gillian_2021}. Thus, users who instantiate the tool only
need to prove the correctness of the implementation of their compiler and
memory models to ensure the correctness of the entire tool. From the perspective of someone trying to instantiate Gillian with
their compiler and memory models, it is essential to understand the underlying
intermediate language GIL and the overall memory model interface used by Gillian.

\subsubsection{GIL}
GIL is the GOTO-based Intermediate Language used by Gillian which is used for
all types of analyses the tool supports. For concrete execution, GIL supports
basic GOTO constructs and assertions. For symbolic execution, the GIL grammar
is extended to support path cutting, i.e. assumptions, and generation of symbolic
variables. For separation logic based verification, the GIL grammar is further
extended to support core predicates and user-defined predicates \cite{maksimovic_2021}
that can be utilised to form separation logic based assertions. Furthermore,
function specifications in the Hoare-triple form $\{P\} f(\bar{x}) \{Q\}$ can
be
provided, where $P$ and $Q$ are separation logic based assertions.

Note that Gillian uses a value set $\mathcal{V}$ which differs from that
used in the CHERI-C memory model. As we are only interested in the values used
in the CHERI-C memory model, it is possible to implement a thin conversion
layer between the two value systems. We note that a list of GIL values
also
constitutes a GIL value, so arguments for functions can be expressed as a single
GIL value. This is important when understanding the memory model layout of Gillian.

\subsubsection{Memory Model}
Memory Models in Gillian have a specific definition and have properties
that state what kind of analysis is supported. Proving that the provided
memory models satisfy certain properties is essential in understanding what the
instantiated tool supports.

Gillian differentiates between concrete and symbolic memory models, which
are used for concrete and symbolic execution, respectively. As we are concerned
with concrete execution, we will consider only concrete memory models here.

At the highest level, there are two kinds of memory model properties:
\emph{executional}
and \emph{compositional}. The \emph{executional} memory model states properties
a memory
model must have for whole-program execution, and the
\emph{compositional} memory model states properties a memory model must have for
separation logic based symbolic verification. Each paper in the Gillian literature states
slightly different definitions for the
memory models \cite{fragoso_2020, old_gillian_2020, maksimovic_2021, gillian_2021}---in Definitions \ref{def:execution_memory_model} and \ref{def:compositional_memory_model} below, we
present unified, consistent definitions for each of the memory model properties.
We ignore contexts, as there exists only one context in concrete memories, which
is the GIL boolean value \texttt{true}.

\begin{definition} (Execution Memory Model).
Given the set of GIL values $\mathcal{V}$ and an action set $A$, an execution
memory model $M(\mathcal{V}, A) \triangleq (|M|, \mathcal{W}_{f}, \underline{ea})$
comprises:
\begin{enumerate}
    \item a set of memories $|M| \ni \mu$
    \item a well-formedness relation $\mathcal{W}_{f} \subseteq |M|$, with
          $\mathcal{W}_{f}(\mu)$ denoting $\mu$ is well-formed
    \item the action execution function $\underline{ea} : A \rightarrow |M|
          \rightarrow \mathcal{V} \rightarrow \mathcal{R}\ (|M| \times \mathcal{V})$
\end{enumerate}
\label{def:execution_memory_model}
\end{definition}

\begin{definition} (Compositional Memory Model).
Given the set of GIL values $V$ and core predicate set $\Gamma$, a compositional memory
model, $M(V, A_\Gamma) \triangleq (|M|,
\mathcal{W}_{f}, \\\underline{ea}_\Gamma)$ comprises:
\begin{enumerate}
    \item a partial commutative monoid (PCM) $(|M|, \cdot, 0)$
    \item A well-formedness relation $\mathcal{W}_{f} \subseteq |M|$ with the
          following property:
          $$
          \mathcal{W}_{f}(\mu_1 \cdot \mu_2) \Longrightarrow
          \mathcal{W}_{f}(\mu_1) \land \mathcal{W}_{f}(\mu_2)
          $$
    \item the predicate action execution function
          $\underline{ea}_{\Gamma} : A_{\Gamma} \rightarrow |M|
          \rightarrow V \rightharpoonup \mathcal{R}\ (|M|
          \times V)$
\end{enumerate}
\label{def:compositional_memory_model}
\end{definition}

First, we note that for concrete execution, Gillian also uses the return type
$\mathcal{R}$ in the action execution function $\underline{ea}$.\footnote{In
the Gillian literature, it is stated that $\mathcal{R}$ can return both a return
value and an error. The OCaml implementation of Gillian slightly differs from
this and is more similar to $\mathcal{R}$ used for the CHERI-C memory model.}
For $\mathcal{W}_f$ defined in Definition
\ref{def:execution_memory_model},
the main properties that must be satisfied are Properties 3.1, 3.2, and 3.6 in
\cite{gillian_2021}.

The PCM requirement is required to show that the heap forms a separation algebra
\cite{calcagno_2007}. $\mathcal{W}_f$ is extended to state that
memory composition must also be well-formed. Finally, the predicate action execution
function $\underline{ea}_{\Gamma}$ provides a way to frame on and off parts
of the memory, though they are not required for concrete execution as they are
 not part of the GIL concrete execution grammar.

Using the CHERI-C memory model we defined earlier, we can show that our model
conforms to both Definitions \ref{def:execution_memory_model} and
\ref{def:compositional_memory_model}.
Let $A_{\mathcal{C}}$ be the set of memory
actions, $\mathcal{H}$ be the memory,
$\underline{ea}_\mathcal{C}$ be the action execution function of the CHERI-C
memory model, and $\mathcal{W}^\mathcal{C}_{f}$ be the well-formedness relation.
Then we observe that $(\mathcal{H}, \mathcal{W}^\mathcal{C}_f,
\underline{ea}_\mathcal{C})$ forms an execution memory model. We note that
Properties 3.1 and 3.2 in \cite{gillian_2021} are satisfied, and Property 3.6
is trivial in that operations that return errors do not return an updated heap.
We also note that the memory model also conforms to a compositional memory
model, as we have the PCM $(\mathcal{H}, \uplus, \mu_0)$ along with the
well-formedness property being composition-compatible. The predicate action
execution function is not required to be given, as the concrete execution of
Gillian does not utilise this feature.

\subsection{Compiler}
We implemented a CHERI-C to GIL compiler by utilising ESBMC's GOTO language.
The idea
is that ESBMC uses its own intermediate representation for bounded model checking,
which is the GOTO language. CHERI-enabled ESBMC uses Clang as a front end to
generate the GOTO language. In our case we can build a GOTO to GIL compiler
instead of building a CHERI-C compiler from scratch. The GOTO language is very
similar to GIL in that they are both goto-based languages and uses single
static assignment. For most parts, the compilation process is straightforward.
As ESBMC's GOTO language is typed while the CHERI-C memory model is untyped---untyped in the sense that the memory model does not support user-defined
types like \texttt{struct}s---we make sure that capability arithmetic and casts
are
applied correctly by inferring the sizes of the user-defined types.
\section{Experimental Results}\label{sec:expt}
In Sect.~\ref{sec:app}, we have provided a way to instantiate the Gillian
tool, where we obtain a concrete CHERI-C model using Isabelle/HOL and a
CHERI-C to GIL compiler that utilises ESBMC's GOTO language. Our framework
can demonstrate that higher-level memory actions---such as
\texttt{memcpy()}, which preserves tags when applicable---can be
implemented. Furthermore, we can run concrete instances of programs that
use \texttt{memcpy()} to show they emit the expected behaviour. This also
means the tool can catch the TagViolation exception that is triggered in
Listing \ref{lst:c_example}. Our tool also allows capability-related
functions defined in \texttt{cheriintrin.h} and \texttt{cheri.h}, to be
usable, i.e. it is possible to call operations such as
\texttt{cheri\_tag\_get()} and \texttt{cheri\_tag\_clear()}.


\begin{table}
\parbox{.55\linewidth}{
\centering
    \begin{tabular}{ |l|c|c|c|c| }
        \hline
        \textbf{Filename} & \textbf{GC} & \textbf{GCC} &
\textbf{AM} & \textbf{BMC}\\
        \hline
        \texttt{buffer\_overflow.c} & $\checkmark$ & $\checkmark$ &
$\checkmark$ & $\checkmark$  \\
        \texttt{dangling\_ptr.c} & $\checkmark$ & $\checkmark$ &
\texttimes  & $\checkmark$ \\
        \texttt{double\_free.c} & $\checkmark$ & $\checkmark$ &
\texttimes  & $\checkmark$ \\
        \texttt{invalid\_free.c} & \texttimes\textsuperscript{{\ref{foot:gil_bug}}} & $\checkmark$ &
$\checkmark$  & $\checkmark$\\
        \texttt{misaligned\_ptr.c} & $\checkmark$ & $\checkmark$ & $\checkmark$
& \texttimes\\
        \texttt{listing\_1.c} & \texttimes & $\checkmark$ &
$\checkmark$ & \texttimes \\
        \hline
    \end{tabular}
\caption{\centering Violation detection\label{tab:compare}}
}
\hfill
\parbox{.35\linewidth}{
\centering
    \begin{tabular}{ |l|r| }
        \hline
        \textbf{Filename} & \textbf{Time}(s) \\
        \hline
        \texttt{libc\_malloc.c} & \textbf{8.585} \\
        \texttt{libc\_memcpy.c} & 1.698 \\
        \texttt{libc\_memmove.c} & 0.318 \\
        \texttt{libc\_string.c} & 0.315 \\
        \hline
    \end{tabular}
\caption{\centering GCC runtime \hspace{\textwidth}performance\label{tab:perf}}
}
\vspace{-2em}
\end{table}
Table \ref{tab:compare} shows a list of safety violations that Gillian-C, our
tool, the ARM Morello hardware, and CHERI-ESBMC---labelled as GC, GCC, AM, and BMC,
respectively---all catch.
We observe that Morello 
fails to catch temporal safety violations such as dangling pointers and double 
frees. For the invalid free case, where we attempt to free a pointer not produced 
by \texttt{malloc}, we discovered a bug in the Gillian-C tool that
fails to catch
this violation.\footnote{\label{foot:gil_bug}The bug has since been fixed after a discussion with the
developers~\cite{gillian_github_commit}.}
Gillian-C does not return any errors
for the program in Listing \ref{lst:c_example}, which is to be expected, as this
is not problematic for conventional C. Finally, we observe
that CHERI-ESBMC fails to catch the last two violations that
relating to tag invalidation.

Table \ref{tab:perf} shows the runtime performance of running the CHERI-C
library test suites, based on the Clang CHERI-C test suite \cite{cheri_c_tests}.
Tests were conducted on a machine running Fedora 34 on
an $11^{\text{th}}$ Gen Intel
Core i7-1185G7 CPU with 31.1 GB RAM, with trace logging enabled.
We note that when the test cases were executed on Morello
without any modifications to the code, all of the tests
terminated instantaneously without any issues. 
In the \texttt{libc\_malloc.c} test case, 
we reduced the scope of the test\footnote{In particular, we reduced \texttt{max} from the \texttt{libc\_malloc.c} case in~\cite{cheri_c_tests} from 20 to 9.}
to ensure the tool terminates within a
reasonable time, though the performance can be drastically improved by 
turning logging off, e.g. the \texttt{libc\_malloc.c} case would only take 
0.686 seconds.
For the remaining tests, we made modifications to the code to ensure the
compiler can correctly produce the GIL code, and we made sure to preserve
all the edge cases covered by the original tests. For example, in
\texttt{libc\_memcpy.c} we made sure to test all cases where both \texttt{src}
and \texttt{dst} capabilities were aligned and misaligned in the beginning and
the end, which affected tag preservation. We observed that no assertions were
violated, and we also observed that the same code when run in Morello also
resulted in no assertion violations, demonstrating a faithful implementation
of CHERI-C semantics. 

%
\section{Related Work}\label{sec:related}

The CompCert C memory model~\cite{leroy_2012}, CH$_2$O memory 
model~\cite{krebbers_2016}, and Tuch's C memory model~\cite{tuch_2009} are C 
memory models formalised in a theorem prover, each focusing on different aspects
of verification. Our model mostly draws inspiration from these
models, extending such work to support CHERI-C programs.

VCC, which internally uses the typed C memory model~\cite{cohen_2009}, and 
CHERI-ESBMC~\cite{brausse_2022} are designed with automated verification of C 
programs via symbolic execution in mind---in particular, CHERI-ESBMC supports 
hybrid settings and compressed capabilities in addition to purecap settings and
uncompressed capabilities. Both tools rely on a memory model that is not 
formally verified, so the tools have components that must be trusted.





%
\section{Conclusion and Future Work}\label{sec:concl}
We have provided a formal CHERI-C memory model and demonstrated its utility
for verification. We formalised the entire
theory in Isabelle/HOL and generated an executable instance of the memory model,
which was then used to instantiate a CHERI-C tool. The result lead to a
concrete execution tool that is robust in terms of the properties that are
guaranteed both by the tool and by the memory model. We demonstrated its
practicality by running CHERI-C based test suites, capturing memory safety
violations, and comparing the results with actual CHERI hardware---namely the physical
Morello processor.

Currently there are a number of limitations provided by the memory model.
Capability arithmetic is limited only to addition and subtraction, but the heap
can be extended to incorporate mappings from blocks to physical addresses
and vice versa. This provides a way to extend capability arithmetic.
While the theory incorporates abstract capabilities, compression is still under
work. We believe, however, that the abstract design itself does not need to change.
It may be possible to utilise the compression/decompression work to convert
between the two forms \cite{cheri_compressed_cap} when needed whilst retaining
our design for the operations.

This theory serves as a starting point for much potential future work. A compositional
symbolic memory model can be built from this design to enable symbolic execution
and verification in Gillian. As we have already proven the core properties,
proving the remaining properties for the extended model will allow automated
separation logic based verification of CHERI-C programs.

\subsubsection*{Acknowledgements}
We are very grateful to the Gillian team, in particular, Sacha{-}{\'{E}}lie 
Ayoun, for providing assistance with instantiating the Gillian tool. We also 
thank Fedor Shmarov and Franz Brau{\ss}e for providing assistance with 
building and modifying the ESBMC tool. This work was funded by the UKRI
programme on Digital Security by Design (Ref. EP/V000225/1, 
SCorCH~\cite{SCorCH}).

\subsubsection*{Data-Availability Statement}
The Isabelle/HOL formalisation of the CHERI-C memory model
described in Sect.~\ref{sub:isabelle} is
available in the Isabelle Archive of Formal Proofs~\cite{park_2022}.
The artefact of the evaluation provided in Sect.~\ref{sec:expt}, which includes
Gillian-CHERI-C itself, CHERI-ESBMC, and other tools, is archived in the
Zenodo open-access repository~\cite{park_artefact_2022}.

\bibliographystyle{splncs04}
\bibliography{Gillian}

\end{document}